\documentclass[journal]{IEEEtran}

\usepackage[utf8]{inputenc} 
\usepackage[T1]{fontenc}    
\usepackage{hyperref}       
\usepackage{url}            
\usepackage{booktabs}       
\usepackage{amsfonts}       
\usepackage{nicefrac}       
\usepackage{microtype}      
\usepackage{lipsum}		
\usepackage{graphicx}
\usepackage{amsmath}
\usepackage{float}
\usepackage{makecell}
\usepackage{multirow}
\usepackage{float}
\usepackage{color}
\begin{document}
	\title{Prior Attention Network for Multi-Lesion Segmentation in Medical Images}
	\author{Xiangyu Zhao, Peng Zhang, Fan Song, Chenbin Ma, Guangda Fan, Yangyang Sun, Youdan Feng, and Guanglei Zhang
		
		\thanks{This work is partially supported by the Fundamental Research Funds for Central Universities, the National Natural Science Foundation of China (No. 61601019, 61871022), the Beijing Natural Science Foundation (No. 7202102), and the 111 Project (No. B13003).}
		\thanks{Corresponding Author: Guanglei Zhang (e-mail: guangleizhang@buaa.edu.cn)}
		\thanks{X. Zhao, P. Zhang, F. Song, C. Ma, G. Fan, Y. Sun, Y. Feng are with the School of Biological Science and Medical Engineering, Beihang University, Beijing, 100191, China (e-mails: \{zhaoxy, pengzhang, fansong, machenbin, a641261717, syyzbh, emilyfeng\}@buaa.edu.cn)}
		\thanks{G. Zhang is with the Beijing Advanced Innovation Center for Biomedical Engineering, School of Biological Science and Medical Engineering, Beihang University, Beijing, 100191, China (e-mail: guangleizhang@buaa.edu.cn)}}
	
	\maketitle
	\bibliographystyle{ieeetr}
	
	\begin{abstract}
		The accurate segmentation of multiple types of lesions from adjacent tissues in medical images is significant in clinical practice. Convolutional neural networks (CNNs) based on the coarse-to-fine strategy have been widely used in this field. However, multi-lesion segmentation remains to be challenging due to the uncertainty in size, contrast, and high interclass similarity of tissues. In addition, the commonly adopted cascaded strategy is rather demanding in terms of hardware, which limits the potential of clinical deployment. To address the problems above, we propose a novel \textit{Prior Attention Network} (\textit{PANet}) that follows the coarse-to-fine strategy to perform multi-lesion segmentation in medical images. The proposed network achieves the two steps of segmentation in a single network by inserting lesion-related spatial attention mechanism in the network. Further, we also propose the intermediate supervision strategy for generating lesion-related attention to acquire the regions of interest (ROIs), which accelerates the convergence and obviously improves the segmentation performance. We have investigated the proposed segmentation framework in two applications: 2D segmentation of multiple lung infections in lung CT slices and 3D segmentation of multiple lesions in brain MRIs. Experimental results show that in both 2D and 3D segmentation tasks our proposed network achieves better performance with less computational cost compared with cascaded networks. The proposed network can be regarded as a universal solution to multi-lesion segmentation in both 2D and 3D tasks. The source code is available at: \url{https://github.com/hsiangyuzhao/PANet}.
	\end{abstract}
	
	
	\section{Introduction}
	\label{sec:introduction}
	Medical image segmentation is significant for the accurate screening of diseases and the prognosis of patients. The evaluation of lesions based on lesion segmentation provides the information about disease progression and help physicians improve the quality of clinical diagnosis and treatment. However, manual lesion segmentation is rather subjective and laborious, which limits its potential clinical application. Recently, with the fast development of artificial intelligence, algorithms based on deep learning have been widely used and have achieved the state-of-the-art performance in medical image segmentation \cite{litjens2017survey}. Convolutional neural networks (CNNs) are popular for lesion segmentation in medical images owing to its high segmentation quality. Such algorithms typically feature a deep encoder to extract features automatically from the input images and the following operations to generate dense predictions. For example, Long et al. \cite{long2015fully} proposed a fully convolutional network for image semantic segmentation, which is rather influential and inspires the later end-to-end frameworks in medical segmentation. Ronneberger et al. \cite{ronneberger2015u} proposed a U-shaped network (U-Net) for medical image segmentation, which has shown promising results in many fields of medical segmentation and has become a virtual benchmark in many medical segmentation tasks. 
	
	However, despite those breakthroughs in medical segmentation, current practices of medical segmentation mainly focus on the binary segmentation of lesions, i.e. distinguishing lesions (foreground) and everything else (background). Although binary segmentation does help improve the clinical decisions of physicians, the information provided by the binary segmentation is still limited. In clinical scenarios, physicians usually need to distinguish different types of lesions, i.e. multi-lesion segmentation. Compared with binary segmentation, such scenarios are much more difficult due to the interclass similarity of the tissues, as different kinds of lesions can resemble in texture, size and shape. Cascaded networks with a coarse-to-fine strategy have been extensively used in such scenarios, such as the segmentation of liver and lesions, and brain tumor segmentation \cite{christ2016automatic} \cite{wang2017automatic} \cite{li2019multi} \cite{jiang2019two}. Such networks usually consist of two independent networks, where the first network performs the coarse segmentation and the second network refines the segmentation based on the ROIs segmented from the first network. However, although cascaded networks have been widely used in multi-lesion segmentation in medical images, the cascade strategy has its own drawbacks. Since the cascaded network consists of two independent networks, the parameter amount and video memory occupation are typically twice those of a single network, which could be demanding in hardware and limits its potential in clinical usage. More importantly, as the two networks in a cascaded network are usually independent, the training process of the cascaded network is sometimes more difficult than a single network, which could lead to underfitting.
	In this paper, we present a novel network structure named \textit{Prior Attention Network} (PANet) to perform multi-lesion segmentation in medical images. The proposed network consists of an encoder for feature extraction, and two decoders to generate lesion region attention and final predictions separately. The network is combined together with attention mechanism. To reduce the parameter size and hardware occupation, we use the deep, semantic-rich features from the network encoder to generate the spatial attention of lesion regions. The feature representations generated by the encoder are then refined by the spatial attention and send to the decoder for final multi-class predictions. To improve segmentation performance and accelerate the convergence, we also introduce intermediate supervision and deep supervision to the network structure. With these improvements, the proposed network achieves competitive results with significantly lower parameter size and computational cost compared with traditional cascaded networks.
	\subsection{Contributions}
	The contributions of this work lies in three aspects. First, we propose a novel network architecture which follows the coarse-to-fine strategy for multi-lesion segmentation in 2D and 3D medical images by combining the two steps of segmentation in traditional cascaded networks in a single network. Compared to cascaded networks, the proposed architecture achieves competitive segmentation performance with much less additional computational cost, which is much easier to train and deploy to production environments. Second, we propose a spatial attention-based method that combines the attention of lesion regions with the features extracted by the network, which enables the efficient synchronous training of different modules in the network. Third, we introduce intermediate supervision and deep supervision strategy into the proposed network to acquire better convergence of the network and improve the performance of lesion segmentation.
	\begin{figure}[ht]
		\centering
		\includegraphics[width=0.44\textwidth]{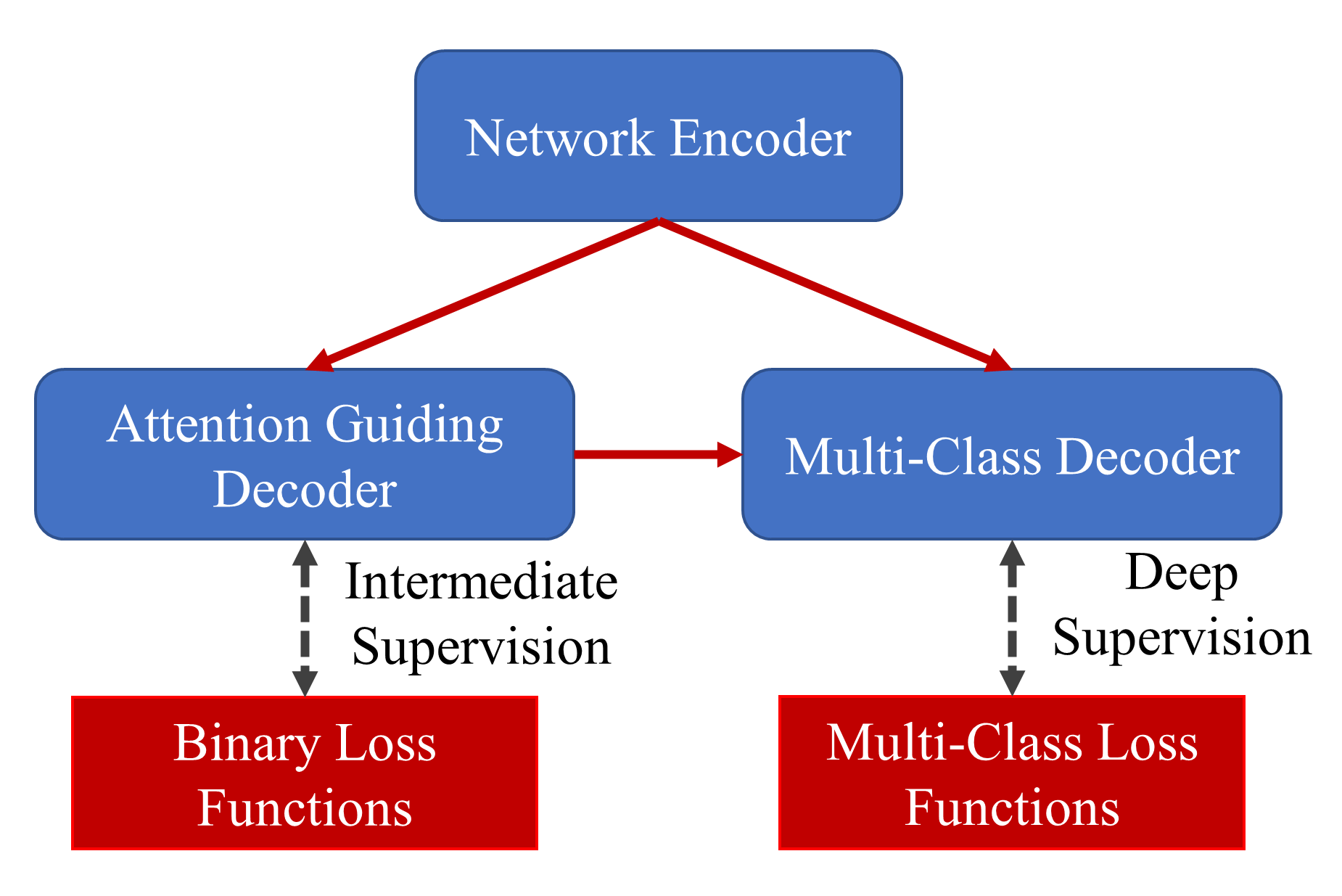} 
		\caption{The basic scheme of the proposed \textit{Prior Attention Network} with intermediate supervision and deep supervision.}
		\label{fig:scheme}
	\end{figure}
	\subsection{Related Works}
	\subsubsection{Network Structures for Image Segmentation}
	A typical convolutional neural network for image segmentation usually consists of a convolutional feature extractor whose topology is similar to common classification networks to extract features from input images automatically and following convolution-based operations to generate final dense predictions. In the field of natural image segmentation, FCN \cite{long2015fully}, DeepLab \cite{chen2017rethinking}, PSPNet \cite{zhao2017pyramid} and SegNet \cite{badrinarayanan2017segnet} are rather popular for their performance and efficiency. For medical segmentation, U-Net \cite{ronneberger2015u} has been rather popular in many tasks, and has been modified into many improved versions such as Attention U-Net \cite{oktay2018attention}, U-Net++ \cite{zhou2018unet++}, V-Net \cite{milletari2016v} and H-DenseUNet \cite{li2018h} to acquire better performance in certain fields.
	\subsubsection{Cascaded Networks in Medical Segmentation}
	Cascaded networks have been widely applied in the segmentation of normal tissues and lesions and the segmentation of different types of lesions, including the segmentation of liver lesions, brain tumors, sclerosis lesions and prostate cancers \cite{christ2016automatic} \cite{valverde2017improving} \cite{wang2017automatic} \cite{zhu2019fully}. For example, Awad et al.\cite{albishri2019cu} presented a cascaded framework called CU-Net to perform automatic segmentation of liver and lesions in CT scans. They also provided useful information and interpretation which could guide clinical treatments. Xi et al. \cite{xi2020cascade} proposed a Cascaded U-ResNets which follows a novel vertical cascade strategy, and different types of loss functions were evaluated in their work. Apart from liver lesion segmentation, the cascade strategy is popular in BraTS challenges as well. For instance, the Top-2 solutions \cite{li2019multi} \cite{jiang2019two} of BraTS 2019 challenge are both cascaded networks with different cascade strategies.
	\subsubsection{Attention in Neural Networks}
	Attention mechanism is inspired by human's perception and visual cognition, and has been commonly applied in computer vision tasks \cite{hu2018squeeze} \cite{woo2018cbam} \cite{wang2017residual} \cite{oktay2018attention}. Attention mechanism in computer vision tasks is to generate the spatial or channel weight maps on feature representations extracted by the neural networks. For instance, Woo et al. \cite{woo2018cbam} developed a convolutional block attention module (CBAM) to introduce a fused attention which includes both channel attention and spatial attention. Such attention module can be inserted to commonly-used classification or segmentation networks. Oktay el al. \cite{oktay2018attention} proposed a novel attention gate in the Attention U-Net, which is utilized to refine the feature representations extracted by the network encoder to facilitate the network to focus on the ROIs. In summary, attention mechanism in neural networks is utilized to highlight the ROIs and suppress irrelevant information, which improves the network capacity and the robustness against noise.
	\begin{figure*}[ht]
		\centering
		\includegraphics[width=0.95\textwidth]{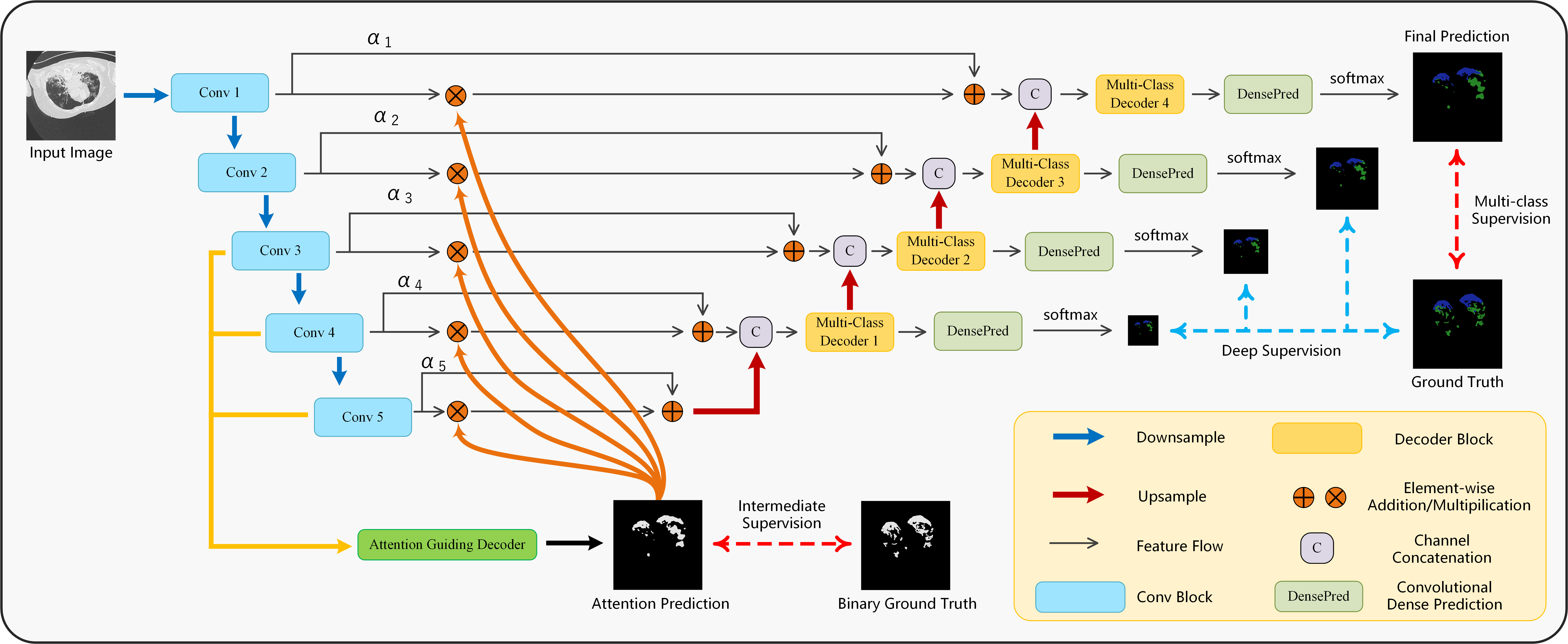} 
		\caption{The topology of the proposed \textit{Prior Attention Network}. We use 2D segmentation of COVID-19 lesions to illustrate the architecture of the proposed network.}
		\label{fig:network}
	\end{figure*}
	
	\section{Methods}
	\label{sec:methods}
	In this section we will go through the details of the proposed \textit{Prior Attention Network} architecture. In the first part, we will offer the overview of proposed network. We then provide details about the proposed \textit{attention guiding decoder} with intermediate supervision, parameterized skip connections and multi-class decoder with deep supervision accordingly.
	\subsection{Overview of Network Architecture}
	\label{sec:overview}
	Basically, our proposed network is modified based on the U-Net \cite{ronneberger2015u} architecture, which features a U-shape topology and the skip connections between the encoder and the decoder. In the proposed \textit{Prior Attention Network}, a novel \textit{attention guiding decoder} module is integrated to the skip connections of the network to refine the feature representations by spatial attention. A novel parameterized skip connection is also introduced to the network to guide the network to learn the ratio between vanilla feature maps and refined feature maps. The \textit{attention guiding decoder} takes the semantic-rich features from the encoder and generate spatial attention maps to guide the following multi-class segmentation. To generate ROI-related attention, the intermediate supervision strategy has been utilized in the framework. The refined feature maps are then sent to the multi-class decoder for final dense predictions. The deep supervision strategy is utilized in the multi-class decoder to acquire better convergence and improve segmentation performance. Such network topology achieves the two steps of conventional cascaded networks in a single network through the attention maps generated by the attention guiding decoder. The network scheme is shown in Fig. \ref{fig:network}.
	\subsection{Attention Guiding Decoder}
	\label{sec:AGD}
	In typical cascaded networks, the first step of the segmentation is to perform a coarse segmentation and find the ROIs in the input images. In the proposed \textit{Prior Attention Network}, we present an \textit{attention guiding decoder} to perform the process. The proposed \textit{attention guiding decoder} is integrated to the network to generated ROI-related attention maps, which is then utilized to refine the feature representations and improve multi-class segmentation performance.
	\subsubsection{Module Topology}
	The basic topology of the proposed \textit{attention guiding decoder} is based on the feature fusion proposed in FCN\cite{long2015fully}. The feature representations extracted from the deepest three layers in the network decoder is fed to the module. As the feature maps differ in spatial size, linear interpolation is first performed to upsample the feature maps. Then the features are compressed to inhibit irrelevant information in the channel dimension and reduce computational cost. The compressed features are then concatenated respectively in the channel dimension for feature fusion. Finally the three feature maps are fused together to acquire the final prediction.
	
	For simplicity, we use 2D segmentation to illustrate the computation of attention maps. We use ${X}_{i} \in \mathbb{R}^{C_i \times H_i \times W_i}, i \in (3,4,5)$ to denote the feature maps extracted from the network encoder, where the $\mathbf{X}_5$ denotes the deepest feature. The feature compression and fusion is computed as follows:
	\begin{equation}
		\mathbf{Z}_{5} = W^T_{c5}\mathbf{X}_{5} \oplus \mathbf{X}_{4}
	\end{equation}
	\begin{equation}
		\mathbf{Z}_{4} = (W^T_{c4}(W^T_{4}\mathbf{Z}_{5})) \oplus \mathbf{X}_{3}
	\end{equation}
	where $\mathbf{Z}_{5} \in \mathbb{R}^{C_{4} \times H_4 \times W_4}$ denotes the fused feature of $\mathbf{X}_{5}$ and $\mathbf{X}_{4}$, $\mathbf{Z}_{4} \in \mathbb{R}^{C_{3} \times H_3 \times W_3}$ denotes the fused feature of $\mathbf{X}_{4}$ and $\mathbf{X}_{3}$, $W_{c5} \in \mathbb{R}^{C_{5} \times C_{4}}$ and $W_{c4} \in \mathbb{R}^{C_{4} \times C_{3}}$ denote the corresponding compression convolutions, $W_{4} \in \mathbb{R}^{C_{4} \times C_{4}}$ denotes the fuse convolutions to fuse $\mathbf{X}_5$ and $\mathbf{X}_4$, and $\oplus$ denotes feature concatenation.
	
	The output of the \textit{attention guiding decoder} is computed as follows:
	\begin{equation}
		\mathbf{Y} = \sigma (W^T_{out}(W^T_3\mathbf{Z}_{4})) \\
	\end{equation}
	where $W_{3} \in \mathbb{R}^{C_{3} \times C_{3}}$ denotes the fuse convolutions to fuse $\mathbf{X}_4$ and $\mathbf{X}_3$, $W_{out} \in \mathbb{R}^{C_{3} \times 1}$ denotes the output convolution, and $\sigma$ denotes \textit{Sigmoid} activation, respectively.
	\begin{figure*}[t]
		\centering
		\includegraphics[width=0.95\textwidth]{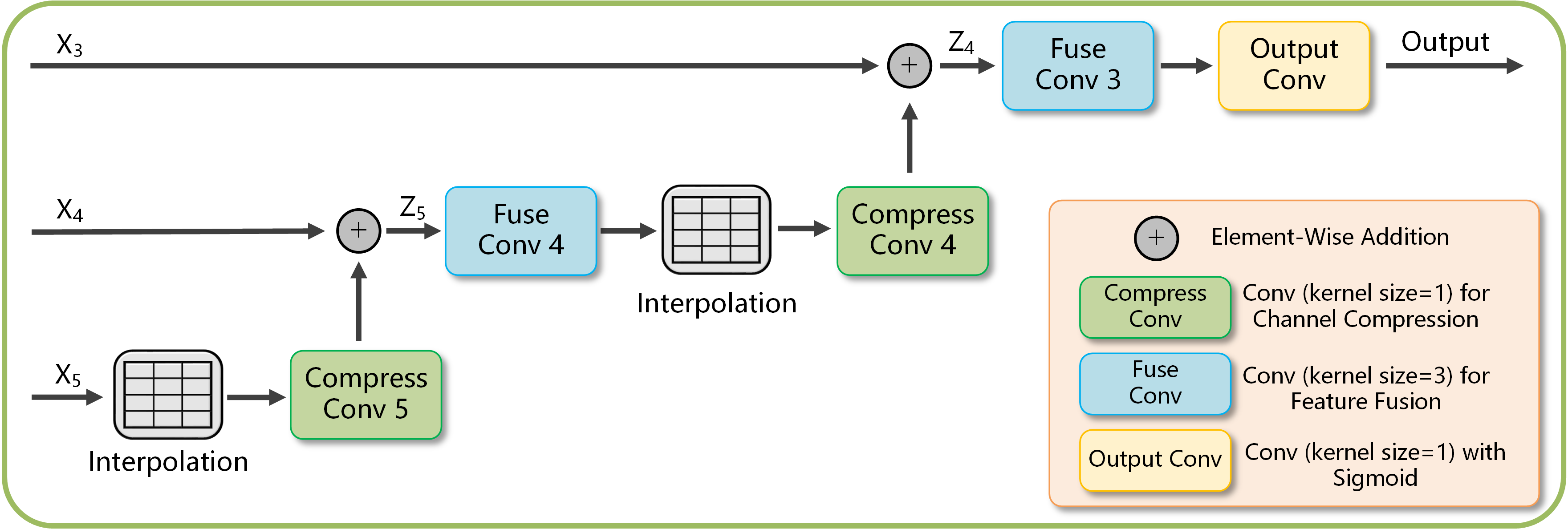} 
		\caption{The topology of the proposed \textit{attention guiding decoder}.} 
		\label{fig:AGD}
	\end{figure*}
	\subsubsection{Intermediate Supervision}
	Traditional attention mechanism in computer vision generates attention maps automatically, but the process of attention generation is usually not humanly interpretable, and the regions that the network focus on may differ from the regions that humans pay attention to. Such gaps can limit the performance and interpretability of the attention mechanism, and sometimes cause the deterioration of network capacity. To address these issues, we introduce an intermediate supervision strategy to the network. In a multi-lesion segmentation task which follows the coarse-to-fine manner, we first generate a binary ground truth where the foreground denotes all types of lesions and the background denotes everything else. In a multi-lesion segmentation task with $C$ types of lesions, we use $\mathbf{G}_i, i \in (1, ..., C)$ to denote the binary ground truth of the $i$-th type of lesion, where the foreground denotes the specific lesion and the background denotes everything else. The binary ground truth $\mathbf{G}_{b}$ is computed as follows:
	\begin{equation}
		\mathbf{G}_{b} = \sum_{i=1}^{C} \mathbf{G}_i
	\end{equation}
	
	Binary loss function is then utilized to compute the binary loss $l$ between the binary ground truth $y_b$ and attention map $\mathbf{Y}$ generated from the \textit{attention guiding decoder}:
	\begin{equation}
		l = \mathcal{L}_{b}(y_{b}, \mathbf{Y})
	\end{equation}
	where $\mathcal{L}_{b}$ denotes the binary loss function.
	
	The computed loss $l$ is then utilized to supervise the parameter update of the \textit{attention guiding decoder}.
	\subsection{Parameterized Skip Connections}
	Skip connections have been widely used in popular convolutional networks including U-Net\cite{ronneberger2015u}, ResNet\cite{he2016deep}, etc. Inspired by \cite{oktay2018attention}, we propose to integrate the attention maps to the skip connections which connect the network encoder and the multi-class decoder. In the skip connections, we also introduce an extra residual path to recover the vanilla feature maps and further improve the segmentation performance. Compared with traditional residual paths, the magnitude factor of the residual path $\alpha_{i}, i \in (1, 2, ..., 5)$ is set as a learnable parameter of the network, which is updated during the back propagation. We believe such settings could add extra non-linearity capacity of the network and enhance the effectiveness of skip connections.
	
	We use $\mathbf{F}_i, i \in (1, 2, ..., 5)$ to denote the vanilla feature maps from the network encoder, and $\mathbf{Y}$ to denote the attention map, the refined feature map $\mathbf{F_r}_i, i \in (1, 2, ..., 5)$ which the multi-class decoder receives is computed as follows:
	\begin{equation}
		\mathbf{F_r}_i = \alpha_{i}\mathbf{F}_i + \mathbf{Y} \cdot \mathbf{F}_i, i \in (1, 2, ..., 5)
	\end{equation}
	
	The refined feature maps are then sent to the multi-class decoder for final multi-class predictions.
	\subsection{Multi-Class Decoder with Deep Supervision}
	The decoder in a U-shaped segmentation network is used to receive the feature maps sent by the encoder, and the segmentation performance refines step by step as the features in the decoder decrease in the number of channels and increase in spatial resolution. However, as the network grows deep, the deepest decoder blocks become hard to train, which could limit the final segmentation performance. Deep supervision strategies have been proposed to train deep convolutional networks\cite{lee2015deeply}\cite{wang2015training}. In the proposed \textit{Prior Attention Network}, the auxiliary predictions are extracted from different levels of decoder blocks and supervised with the same ground truth. 
	
	We use $\mathbf{P}_i, i \in (1, 2, 3)$ to denote the auxiliary predictions from the multi-class decoder, $\mathbf{P}_{m}$ to denote the final multi-class prediction, $g$ to denote the ground truth, and $\mathcal{L}_{m}$ to denote the multi-class loss function. The final multi-class loss is calculated as follows:
	\begin{equation}
		l = \mathcal{L}_{m}(g, \mathbf{P}_m) + \sum_{i=1}^3 \mathcal{L}_{m}(g, \mathbf{P}_i)
	\end{equation}
	
	\section{Experiments}
	We validate the proposed framework with two applications: 2D segmentation of multiple lung infections from lung CT slices (COVID-19 CT Segmentation Dataset\cite{covid19ctseg}) and 3D segmentation of multiple lesions from multi-modality brain MRIs (BraTS 2020 Challenge\cite{menze2014multimodal}).
	\begin{table*}
		\centering
		\caption{Quantitative analysis of lung infection segmentation performance on COVID-19 CT Segmentation Dataset. The best two results are shown in \textcolor{red}{red} and \textcolor{blue}{blue} fonts.}
		\label{table:2dquantitative}
		\setlength{\tabcolsep}{2mm}{
			\begin{tabular}{cccccccccc}
				\toprule
				\multirow{2}{*}{Model}  & \multirow{2}{*}{GFlops} & \multirow{2}{*}{Param.} & \multicolumn{2}{c}{Dice} & \multicolumn{2}{c}{Precision} & \multicolumn{2}{c}{Recall} & \\ 
				\cline{4-9}
				&         &             &  GGO  &   CON.      &  GGO  &   CON.      &  GGO  &   CON.  \\
				\midrule
				U-Net                 &65.11    &    38.33M    & 0.638  & 0.493      & 0.657  & 0.763      & 0.642  & 0.437\\ 
				Attention U-Net       &67.88    &    42.49M    & 0.639  & 0.533      & 0.653  & 0.778      & 0.645  & 0.471\\ 
				Cascaded U-Net        &130.21   &   76.65M     &\textcolor{red}{\textbf{0.656}}&\textcolor{blue}{\textbf{0.578}}&\textcolor{blue}{\textbf{0.669}}&\textcolor{red}{\textbf{0.791}} &\textcolor{red}{\textbf{0.667}}&\textcolor{blue}{\textbf{0.515}}   \\
				\midrule
				\textit{PANet}        &88.77    &   52.76M     &\textcolor{blue}{\textbf{0.645}}&\textcolor{red}{\textbf{0.589}}&\textcolor{red}{\textbf{0.682}}&\textcolor{blue}{\textbf{0.760}} &\textcolor{blue}{\textbf{0.650}}&\textcolor{red}{\textbf{0.544}}   \\
				\bottomrule
		\end{tabular}}
	\end{table*}
	\begin{table*}[h]
		\caption{Ablation analysis of proposed \textit{Prior Attention Network} on COVID-19 CT Segmentation Dataset. DS denotes the deep supervision, IS denotes the intermediate supervision, and AGD denotes the attention guiding decoder.}
		\centering
		\label{table:2dablation}
		\setlength{\tabcolsep}{2mm}{
			\begin{tabular}{lcccccccccc}
				\toprule
				\multirow{2}*{Model}              & \multicolumn{2}{c}{Dice} & \multicolumn{2}{c}{Precision} & \multicolumn{2}{c}{Recall} &  \\
				\cline{2-7}
				&  GGO   &  CON.      &  GGO  &  CON.       &  GGO  &  CON.     \\
				\midrule
				(No.1) Enhanced U-Net             & 0.638        & 0.493        & 0.657        & 0.763       & 0.642  & 0.437   \\
				(No.2) Enhanced + DS              &     0.644    &   0.512      & 0.654        &\textbf{0.803}& 0.635  & 0.442\\
				(No.3) Enhanced + DS + AGD w/o IS &     0.641    &   0.582      &\textbf{0.711}&0.768        & 0.602  & 0.513 \\
				\midrule
				(No.4) \textit{Prior Attention Network}  &\textbf{0.645}&\textbf{0.589}&0.682         &0.760       &\textbf{0.650}&\textbf{0.544}\\ 
				\bottomrule
		\end{tabular}}
	\end{table*}
	\subsection{Strong Baselines and Evaluation Metrics}
	To investigate the performance difference of the network architectures, we compare the proposed \textit{Prior Attention Network} with the most popular methods in medical segmentation, including U-Net\cite{ronneberger2015u}, Attention U-Net\cite{oktay2018attention} and cascaded U-Net, in both 2D and 3D segmentation tasks. It should be noted that, compared with the original version proposed in their own papers, the baseline methods are modified and optimized according to certain tasks in terms of network topology to acquire performance improvement. We introduce residual connections\cite{he2016deep}, batch normalization\cite{ioffe2015batch} and pretrained encoder from ImageNet to the baseline methods for 2D COVID-lesion segmentation task, and we also introduce residual connections, instance normalization and PReLU activations to 3D brain tumor segmentation task. Apart from network topology, the baseline methods share the same data augmentation and training configurations with the proposed \textit{Prior Attention Network}.
	
	For 2D segmentation of COVID-19 lesions, we use Dice index, precision score and recall score to evaluate the performance of the proposed network. Dice index is a statistic used to gauge the similarity of two samples, which has been widely used for the evaluation of segmentation algorithms. Precision score measures the proportion of positive identifications which are actually correct, and recall score measures the algorithm's sensitivity to positive samples. For 3D segmentation of the BraTS 2020 challenge, the evaluation is performed on the online portal and the algorithms are ranked according to the Dice index and 95$\%$ Hausdorff distance (HD).
	
	We use $\mathbf{G}$ to denote ground truth, $\mathbf{P}$ to denote dense predications, $TP$ to denote true positive, $FP$ to denote false positive, $TN$ to denote true negative, and $FN$ to denote false negative. These metrics are calculated as follows:
	
	\begin{equation}
		\begin{split}
			Dice &= \frac{2 |\mathbf{G} \bigcap \mathbf{P}|}{|\mathbf{G}| + |\mathbf{P}|} \\
			&= \frac{2TP}{2TP + FP + FN}
		\end{split}
	\end{equation}
	
	\begin{equation}
		\begin{split}
			Precision &= \frac{TP}{TP + FP}
		\end{split}
	\end{equation}
	
	\begin{equation}
		\begin{split}
			Recall &= \frac{TP}{TP + FN}
		\end{split}
	\end{equation}
	
	\begin{equation}
		HD = \max \{ \sup_{p \in \mathbf{P}} \inf_{g \in \mathbf{G}} d(p, g), \sup_{g \in \mathbf{G}} \inf_{p \in \mathbf{P}} d(p, g) \}
	\end{equation}
	\subsection{2D Multi-Lesion Segmentation of COVID-19 Lesions From Lung CT Slices}
	\begin{figure*}[t]
		\centering
		\includegraphics[width=0.95\textwidth]{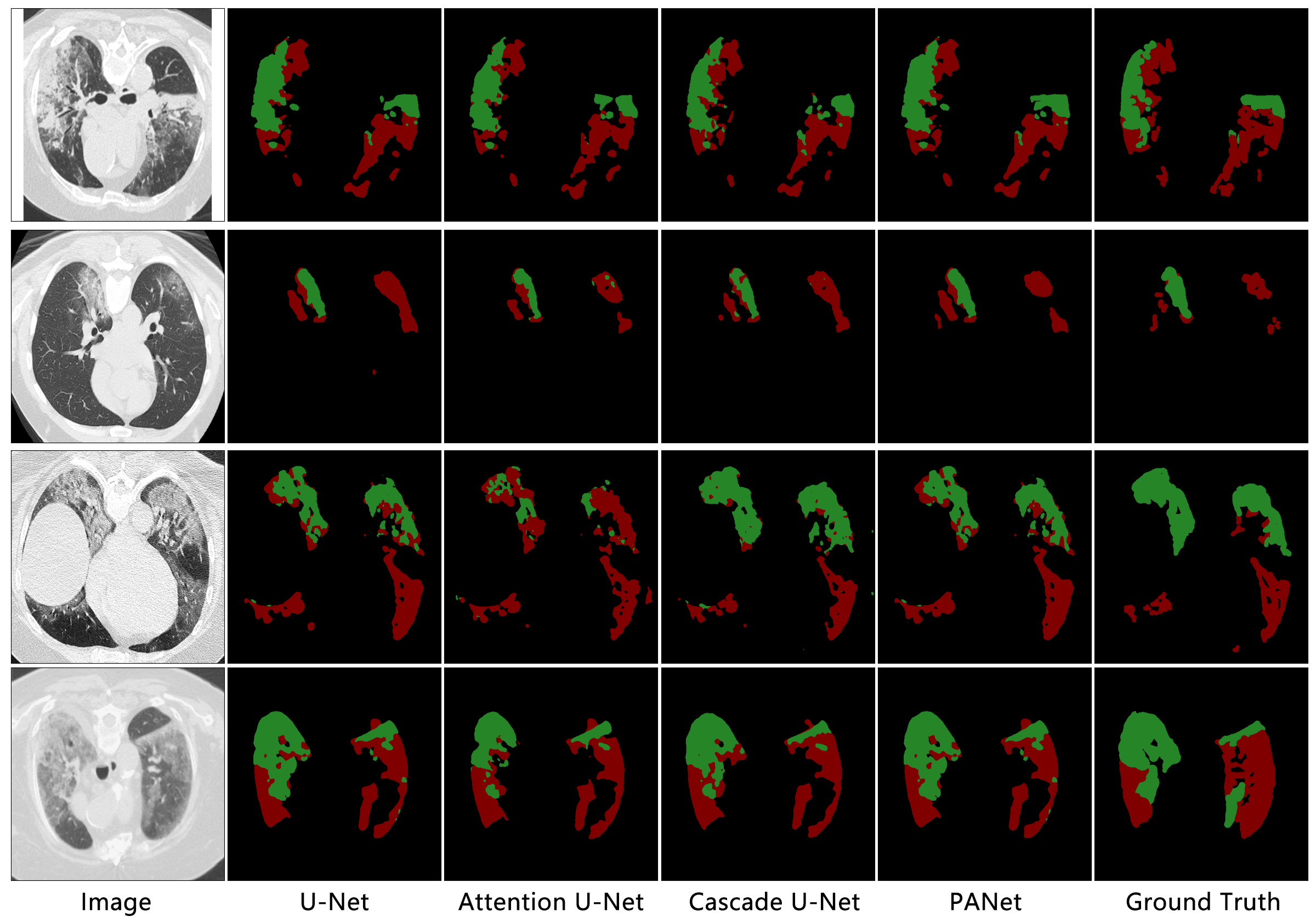} 
		\caption{Visual comparison of the segmentation performance of different models on COVID-19 CT segmentation dataset. The red mask denotes the ground glass opacity and the green mask denotes the consolidation.}
		\label{fig:vis2d}
	\end{figure*}
	\subsubsection{Data}
	We use an open-source COVID-19 CT segmentation dataset \cite{covid19ctseg} in the experiment. The dataset contains 100 axial CT slices from more than 40 patients, which have been rescaled to 512 $\times$ 512 pixels and grayscaled. All slices are segmented by a radiologist with different labels to identify different types of lung infections. We then use 15 randomly selected slices as the test set, and the remaining 85 slices for training. The network is first trained and cross-validated on the training set to find the optimal hyperparameters, and then the network is evaluated on the test set to obtain final segmentation results. The final label and segmentation maps contain 3 classes, including the background, ground glass opacity (GGO) and consolidation (CON.).
	\subsubsection{Implementation Details}
	\paragraph{Model Settings and Loss Functions}
	For pretrained network encoder, we adopt pretrained ResNeXt-50 (32 $\times$ 4d) \cite{xie2017aggregated} from the ImageNet as the encoder for the baseline methods and the proposed \textit{Prior Attention Network}. For upsamling in the decoder, bilinear interpolation is utilized with a scale factor being 2. For the binary loss function in the intermediate supervision and the first stage of the cascaded U-Net, we adopt the linear combination of Dice Loss\cite{milletari2016v} and Focal Loss\cite{lin2017focal} as the loss function. For the multi-class loss function of the final output, we adopt the Focal Tversky Loss\cite{abraham2019novel} as the loss function.
	\paragraph{Training Details}
	Our model is implemented using the PyTorch 1.7.1 framework on an Ubuntu 16.04 server. We use a NVIDIA RTX 2080 Ti GPU to accelerate our training process. Data augmentation is performed with Albumentations\cite{info11020125} in our training process to reduce overfitting and improve generalization capacity. First all input images are rescaled to 560 $\times$ 560, followed by random bright and contrast shift and random affine transform. Then the images are randomly cropped to 512 $\times$ 512, followed by a random elastic transform and finally fed into the network. The model is optimized by an Adam optimizer with $\beta_1 = 0.9$, $\beta_2 = 0.999$, $\epsilon = 1e-8$. $L_2$ regularization is utilized to reduce overfitting as well. We set model weight decay to $1e-5$. Initial learning rate is set to $1e-4$ and reduced followed by cosine annealing strategy. The batch size is set to 4 and the model is trained for 40 epochs.
	\subsubsection{Quantitative Results}
	Detailed comparison among different models in our experiments is shown in Table \ref{table:2dquantitative}. As has been shown, our proposed network outperforms U-Net, Attention U-Net in terms of the Dice score of ground glass opacity and consolidation. And the proposed \textit{Prior Attention Network} achieves competitive results as cascaded U-Net with much less parameters and computational cost. As these models are identical in model backbone and training strategies, it is clear that the combination of proposed \textit{attention guiding decoder}, intermediate supervision and deep supervision contributes to the segmentation performance a lot. The utilization of \textit{attention guiding decoder} aids the model to detect infected tissues more accurately and generate infection-related attention maps, which thus facilitates the multi-class segmentation in the decoder. Besides, the introduction of intermediate supervision and deep supervision promotes the convergence of the network, which contributes to the performance as well.
	\subsubsection{Qualitative Results}
	Visual comparison of different models on 2D COVID-19 slices are shown in Fig. \ref{fig:vis2d}. As the models are quite close in terms of Dice score, these models perform similarly at first glance. But compared with U-Net and Attention U-Net, the proposed \textit{Prior Attention Network} performs better on the segmentation of consolidation and tiny lesions. \textit{Prior Attention Network} produces more accurate segmentation masks compared with U-Net and Attention U-Net, and the proposed network achieves competitive results with much less computational cost compared with Cascaded U-Net. 
	\begin{table*}
		\centering
		\caption{Quantitative analysis of brain tumor segmentation performance on BraTS 2020 Validation Dataset. The best two results are shown in \textcolor{red}{red} and \textcolor{blue}{blue} fonts.}
		\label{table:3dquantitative}
		\setlength{\tabcolsep}{2.2mm}{
			\begin{tabular}{ccccccccccc}
				\toprule
				\multirow{2}*{Model}  & \multirow{2}*{GFlops} & \multirow{2}*{Param.}  & \multicolumn{3}{c}{Dice} & \multicolumn{3}{c}{HD95} \\ 
				\cline{4-9}
				&                       &                        &   ET   &  TC    &    WT  &   ET   &   TC   &   WT     \\
				\midrule
				U-Net                 &586.71                & 16.90M                   & 0.741  & 0.793  & 0.890  &  39.20 &  13.58 & 6.15 \\ 
				Attention U-Net       &588.26                  &    17.12M              & 0.759  & 0.796  & 0.905  &  33.10 &  13.21 & 5.45\\ 
				Cascaded U-Net        &1173.22                 &    33.80M              & 0.736  & 0.810  & 0.908  &  39.23 &  10.41 & 5.24\\
				TransBTS \cite{wang2021transbts} &                 &                   &\textcolor{red}{\textbf{0.787}}& 0.817  & \textcolor{red}{\textbf{0.909}}  &  \textcolor{red}{\textbf{17.95}} &  9.77 & 4.96\\
				Modality-Pairing Learning \cite{wang2020modality}&    &                &\textcolor{blue}{\textbf{0.786}}&\textcolor{red}{\textbf{0.837}}&0.907&  32.25 &\textcolor{blue}{\textbf{8.34}}  & \textcolor{blue}{\textbf{4.39}}\\
				\midrule
				\textit{PANet}        &609.98                  &    19.23M              & 0.784  &\textcolor{blue}{\textbf{0.831}}& \textcolor{red}{\textbf{0.909}} &\textcolor{blue}{\textbf{26.35}} &  \textcolor{red}{\textbf{6.51}} & \textcolor{red}{\textbf{4.02}}\\
				\bottomrule
		\end{tabular}}
	\end{table*}
	\begin{table*}[h]
		\caption{Ablation analysis of proposed \textit{Prior Attention Network} on BraTS 2020 Validation Dataset. DS denotes the deep supervision, IS denotes the intermediate supervision, and AGD denotes the attention guiding decoder.}
		\centering
		\label{table:3dablation}
		\setlength{\tabcolsep}{2mm}{
			\begin{tabular}{lccccccc}
				\toprule
				\multirow{2}*{Model}         & \multicolumn{3}{c}{Dice} & \multicolumn{3}{c}{HD95} \\
				\cline{2-7}
				&   ET   &  TC    &    WT  &   ET   &   TC   &   WT     \\
				\midrule
				(No.1) Enhanced U-Net        & 0.741  & 0.793  & 0.890  &  39.20 &  13.58 & 6.15 \\
				(No.2) Enhanced + DS         & 0.768  & 0.809  & 0.907  &  30.12 &  9.39  & 5.18 \\
				(No.3) Enhanced + DS + AGD w/o IS & \textbf{0.787}  & 0.823  & 0.905  &  \textbf{23.61} &  9.25  & 5.78 \\
				\midrule
				(No.4) \textit{Prior Attention Network}  & 0.784  & \textbf{0.831}  & \textbf{0.909}  &  26.35 &  \textbf{6.51}  & \textbf{4.02} \\ 
				\bottomrule
		\end{tabular}}
	\end{table*}
	\subsubsection{Ablation Analysis}
	Several ablation experiments are conducted to evaluate the performance of the components presented in our model, as shown in Table. \ref{table:2dablation}. 
	\paragraph{Effectiveness of Multi-Class Decoder with Deep Supervision}
	To explore the contribution of deep supervision strategy, we build two baselines: No. 1 (Enhanced U-Net) and No. 2 (Enhanced U-Net + DS). The results in Table. \ref{table:2dablation} reveal that deep supervision has in some degree contributed to the performance.
	\paragraph{Effectiveness of Attention Guiding Decoder}
	We investigate the effectiveness of the proposed \textit{attention guiding decoder} in the proposed network by building baseline No. 3 (Enhanced U-Net + DS + AGD w/o IS). As shown in Table. \ref{table:2dablation}, the introduction of \textit{attention guiding decoder} provides significant performance boost compared with baseline No. 2. This suggests that the \textit{attention guiding decoder} provides effective attention maps in the proposed network, which guides the multi-class segmentation in the decoder.
	\paragraph{Effectiveness of Intermediate Supervision}
	To investigate the effectiveness of the intermediate supervision strategy in the proposed \textit{Prior Attention Network}, we compare the complete version of the proposed network (No. 3) and baseline No. 3. As shown in Table. \ref{table:2dablation}, the proposed network acquires extra improvement compared with baseline No. 3, especially in the segmentation of ground glass opacity.
	\subsection{3D Multi-Lesion Segmentation of Brain Tumor From Multi-Modality Brain MRIs}
	\begin{figure*}[t]
		\centering
		\includegraphics[width=0.95\textwidth]{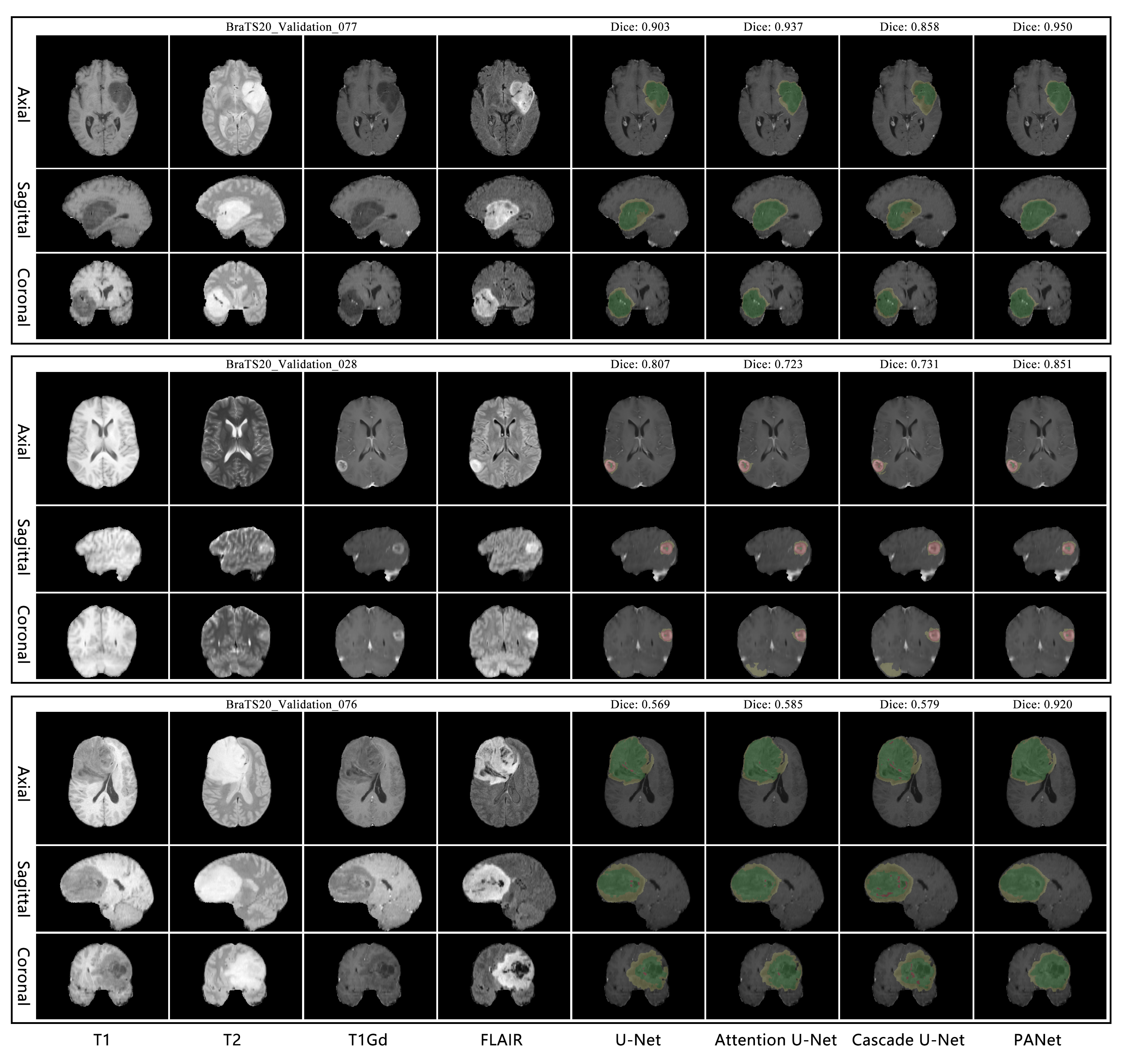} 
		\caption{Visual comparison of the segmentation performance of different models on BraTS 2020 validation dataset. The red mask denotes ET (label 4), the green mask denotes NCR/NET (label 1) and the yellow mask denotes ED (label 2).} 
		\label{fig:vis3d}
	\end{figure*}
	\subsubsection{Data}
	We use the open-source multi-modality MRI dataset from the BraTS 2020 challenge\cite{menze2014multimodal} \cite{bakas2017advancing} \cite{bakas2018identifying}. The training set consists of 369 multi-contrast MRI scans, where each scan contains four modalities, namely native T1-weighted, post-contrast T1-weighted (T1Gd), T2-weighted (T2), and T2 Fluid Attenuated Inversion Recovery (FLAIR). Each scan has a corresponding label with 4 classes: background (label 0), the GD-enhancing tumor (ET, label 4), the peritumoral edema (ED, label 2), and the necrotic and non-enhancing tumor core (NET/NCR, label 1). The validation set consists of 125 multi-contrast MRI scans with the same modalities as the training set and hidden ground truths. All the MRI scans are skull-stripped, aligned to the same brain template (SRI24) and interpolated to 1$mm^3$ resolution. The validation phase is performed via an online portal and the algorithm is ranked due to the performance of 3 overlap tumor regions, i.e. the enhancing tumor (ET), the tumor core (ET + NET/NCR) and the whole tumor (ET + NET/NCR + ED).
	\subsubsection{Implementation Details}
	\paragraph{Model Settings and Loss Functions}
	For 3D segmentation, all the models are trained from scratch due to the lack of open-source pretrained encoders. Downsampling is performed with strided $3 \times 3 \times 3$ padded convolutions, and upsampling is implemented with trilinear interpolation. To further improve the performance on BraTS dataset, we adopt region-based training strategy (optimizing directly on the overlap regions rather than independent labels) and enhancing tumor suppression (replacing predicted enhancing tumor with necrosis if the predicted volume of the enhancing tumor is smaller than a certain threshold) in the training process. For the loss function, we adopt the linear combination of Dice Loss\cite{milletari2016v} and Cross Entropy Loss as the loss function in the both binary and multi-class phase in the network.
	\paragraph{Training Details}
	Our model is implemented using the PyTorch 1.7.1 framework on Ubuntu servers. Since the training of 3D segmentation, especially the cascaded U-Net is rather demanding in video memory, we use a NVIDIA RTX 2080 Ti GPU and a NVIDIA RTX 3090 GPU to train the single model and the cascaded model respectively. Since the video memory occupation of the training process is larger than 11 Gb, we adopt the native mixed precision training procedure provided by the PyTorch framework to save video memory usage and accelerate the training process. The Medical Open Network for AI (MONAI) project \cite{monai} and TorchIO \cite{perez-garcia_torchio_2021} are utilized for the data loading procedures in the training and inference phase respectively. Data augmentation is performed with project MONAI in the training process. First all the modalities are normalized using the z-score normalization separately. Then the images are augmented with random flip, random intensity shift, random intensity scale and elastic transform. Finally we randomly crop the image patch into 128 $\times$ 128 $\times$ 128 and feed it to the network. For inference, we also adopt patch-based inference pipeline to generate predictions on the BraTS 2020 validation set. The patch size is set to 128 $\times$ 128 $\times$ 128 and the overlap between patches is set to 75\%. The predictions in the overlap area are the average of overlapping patches. Such overlap configuration can be regarded as a self-ensemble and produce better segmentation performance.
	\subsubsection{Quantitative Results}
	Detailed comparison among different models in our experiments is shown in Table \ref{table:3dquantitative}. Apart from commonly-used networks based on the U-Net, we also compare the proposed network with the Top-2 solution in the BraTS 2020 \cite{wang2020modality} and the transformer TransBTS in the study. Our proposed \textit{Prior Attention Network} outperforms U-Net, Attention U-Net, cascaded U-Net and TransBTS on the BraTS 2020 validation set. Also, the proposed \textit{Prior Attention Network} achieves similar Dice score with the Top-2 solution on the BraTS 2020 with much better Hausdorff distance. Also, it should be noted that the proposed \textit{Prior Attention Network} only adds to 3.9 \% extra GFlops compared with U-Net, and achieves much better performance than cascaded U-Net with much less computational cost. Such results indicate that the proposed \text{Prior Attention Network} acquires a sophisticated balance between segmentation performance and computational efficiency on the BraTS 2020 dataset.
	\subsubsection{Qualitative Results}
	We visualize 3 MRI images with different segmentation difficulty to demonstrate the performance of different models. For the simplest case (BraTS20\_Validation\_077), all of the models are able to segment the lesions while the proposed \textit{Prior Attention Network} obtains the highest segmentation Dice score. For the case with certain difficulties (BraTS20\_Validation\_028), both Attention U-Net and cascaded U-Net produce serious false positive in terms of the segmentation of edema, which leads to the low Dice score of whole tumor. For the hardest case (BraTS20\_Validation\_076), due to the false positive segmentation of enhancing tumor, the performance of U-Net, Attention U-Net and cascaded U-Net is not promising, while the proposed \textit{Prior Attention Network} yields the best segmentation performance. The success of the proposed \textit{Prior Attention Network} is owed to the \textit{attention guiding decoder} with intermediate supervision, the feature maps are refined by the spatial attention maps, which could prohibit potential false positive predictions.
	\subsubsection{Ablation Analysis}
	Similar ablation experiments as 2D segmentation are also conducted on the BraTS 2020 dataset to evaluate the effectiveness of the components presented in our model, as shown in Table. \ref{table:3dablation}. 
	\paragraph{Effectiveness of Multi-Class Decoder with Deep Supervision}
	We build baseline No. 2 (Enhanced + DS) by introducing deep supervision to the decoder of the network. Compared with the vanilla baseline (No.1), the baseline No.2 to a certain degree provides performance boost in the segmentation performance of enhancing tumor, tumor core and whole tumor.
	\paragraph{Effectiveness of Attention Guiding Decoder}
	We build baseline No. 3 (Enhanced U-Net + DS + AGD w/o IS) to investigate the effectiveness of \textit{attention guiding decoder} in the proposed network. Compared with No. 2, \textit{attention guiding decoder} produces significant improvement in the segmentation performance of enhancing tumor and tumor core. This suggests that \textit{attention guiding decoder} produces effective attention on the feature maps extracted from the encoder, which makes it easier to distinguish the lesions.
	\paragraph{Effectiveness of Intermediate Supervision}
	We investigate the effectiveness of intermediate supervision strategy by the comparison of No. 3 and No. 4 in Table. \ref{table:3dablation}. In No. 4 (Enhanced + DS + AGD w/ IS), by introducing the intermediate supervision, the Dice score of tumor core and whole tumor increases slightly, and the Hausdorff distance of tumor core and whole tumor is improved significantly, compared with No. 3.
	
	\section{Discussion and Conclusion}
	\label{sec:conclusion}
	Multi-lesion segmentation has great significance in the clinical scenarios, as multiple types of infections may occur simultaneously in a certain disease, and patients at different infection stages can have different types of lesions. For instance, ground glass opacity (GGO) and consolidation (CON.) are typical lung lesions in COVID-19 patients, where the former usually happens to early patients, and the increase of the latter may indicate the deterioration of the condition. Gliomas can be classified into low grade gliomas (LGG) and glioblastoma, i.e. high grade gliomas (GBM/HGG), and enhancing tumor is more likely to be found in those who develop HGG. Thus, multi-lesion segmentation has tremendous potential in the screening and prognosis of the patients.
	
	Multi-lesion segmentation problems can be decomposed into a coarse segmentation where the lesions are segmented roughly and a following fine segmentation based on the former one to produce final segmentation maps. Cascaded networks are widely used in multi-lesion segmentation tasks, as the logic behind these algorithms is quite natural. However, the cascaded networks are limited in potential clinical deployments, as they lack flexibility and can be quite demanding in terms of computational resources. In contrast to the existing cascaded networks, we develop a \textit{Prior Attention Network} which integrates both coarse and fine segmentation into one single network. The advantage of our proposed network architecture is the balance of the segmentation performance and computational efficiency. By combining the two steps of segmentation in a single network, the proposed \textit{Prior Attention Network} enables the end-to-end training in multi-lesion segmentation, which has more flexibility in both training and inference and has more potential in clinical deployment.
	
	In conclusion, we have proposed a novel segmentation network, \textit{Prior Attention Network}, for multi-lesion segmentation in medical images. Inspired by the popular coarse-to-fine strategy, we aggregate the two steps of a cascaded network into one single net through spatial attention mechanism. Besides, we introduce a novel intermediate supervision mechanism to guide the generation of lesion-related attention maps, which could guide the following multi-class segmentation in the decoder. The proposed network is evaluated on both 2D and 3D medical images, including CT scans and multi-modality MRIs. For 2D segmentation, the proposed \textit{Prior Attention Network} obtains competitive results with much less computational cost compared with cascaded U-Net. For 3D brain tumor segmentation, the proposed \textit{Prior Attention Network} yields the state-of-the-art performance on the BraTS 2020 validation dataset and outperforms other baseline methods based on U-Net. The experimental results reveal that the proposed method has great potential to be applied in many multi-lesion segmentation tasks in medical imaging, which could provide further information compared with binary segmentation and aid physician's clinical diagnosis in the future.
	
	\bibliography{references}
	
\end{document}